\documentclass{emulateapj}
\usepackage{graphicx}

\newcommand{\E}[1]{\times 10^{#1}}

      \newcommand{\ps}{\,{\rm s}^{-1}}
\newcommand{\yr}{\,{\rm yr}}    \newcommand{\Msun}{M_{\odot}}
\newcommand{\cm}{\,{\rm cm}}    \newcommand{\km}{\,{\rm km}}

\newcommand{\parsec}{\,{\rm pc}} \newcommand{\pc}{\,{\rm pc}}
        \newcommand{\K}{\,{\rm K}}

\newcommand{\XMM}{{\sl XMM}}

\newcommand{\Rb}{R_{\rm b}} \newcommand{\pu}{p_5}

\newcommand{\vw}{v_{\rm w}}

\shorttitle{}

\begin{document}

\title{Linear relation for wind-blown bubble sizes of main-sequence OB stars
in a molecular environment and implication for supernova progenitors}

\author{
 Yang Chen\altaffilmark{1,2},
 Ping Zhou\altaffilmark{1,4},
 You-Hua Chu\altaffilmark{3}
}
\altaffiltext{1}{Department of Astronomy, Nanjing University, Nanjing~210093,
       China}
\altaffiltext{2}{Key Laboratory of Modern Astronomy and Astrophysics,
 Nanjing University, Ministry of Education, Nanjing~210093, China}
\altaffiltext{3}{Department of Astronomy, University of Illinois at
       Urbana-Champaign, 1002 West Green Street, Urbana, IL 61801, USA}
\altaffiltext{4}{Department of Physics and Astronomy, University of Manitoba,
       Winnipeg, MB R3T 2N2, Canada}

\begin{abstract}

We find a linear relationship between the size of a massive star's
main-sequence bubble in a molecular environment and the star's initial
mass: $\Rb\approx1.22M/\Msun-9.16\parsec$,
assuming a constant interclump pressure.
Since stars in the mass range of 8 to 25--$30\Msun$ will end their evolution
in the red supergiant phase without launching a Wolf-Rayet wind,
the main-sequence wind-blown bubbles are mainly responsible
for the extent of molecular gas cavities, while the effect of
the photoionization is comparatively small.
This linear relation can thus be used to infer the masses of
the massive star progenitors
of supernova remnants (SNRs) that are discovered to evolve in molecular cavities,
while few other means are available for inferring properties of SNR progenitors.
We have used this method to estimate the initial masses of the progenitors
of eight SNRs: Kes~69, Kes~75, Kes~78, 3C~396, 3C~397, HC~40, Vela,
and RX~J1713$-$3946.

\end{abstract}


\section{Introduction}


An isolated massive star modifies its surroundings with various
stellar winds throughout its evolution.  During the main sequence
stage, the fast stellar wind sweeps up the ambient interstellar medium
to form an interstellar bubble. 
During the red supergiant (RSG) or luminous blue variable (LBV) phase,
the copious slow wind builds up a circumstellar nebula, which is subsequently
swept into a circumstellar bubble by the fast wind if the Wolf-Rayet
(WR) phase follows 
\citep[e.g.,][]{Getal96a,Getal96b}.
As a massive star ends its life in an explosion and forms a supernova
remnant (SNR), the SNR shocks rapidly pass through the circumstellar
material and impact the massive shell of the interstellar bubble
with drastic deceleration \citep{CZWW03},
and such SNRs
reflect mostly the interstellar bubble sizes.

Massive stars are born in molecular clouds, and the clouds are
most likely clumpy.
In a giant molecular cloud, the mean pressure is $p/k\sim10^5\cm^{-3}\K$
\citep[e.g.,][]{Krumholz_etal2009}, and
this number is indeed needed to confine the dense clumps and to support
the cloud against gravitational collapse \citep{Bl93,Ch99}.
In such an environment,
a supernova (SN) will explode in an interclump medium rather than in a
dense clump, as the SN progenitor has blown a bubble or
carved a cavity via its energetic stellar winds.  The final size
of the stellar wind bubble created in a molecular environment during
the main-sequence stage depends on the stellar mass loss, 
wind velocity, and environmental pressure.  By expressing these
parameters as functions of stellar mass, 
we find a nearly
linear relationship between the bubble size and the initial mass of
the star. This relationship provides a powerful way to assess the
initial masses of the progenitors for SNRs associated with molecular
clouds.



\begin{center}
\begin{deluxetable*}{ccccccccc|c}
\tabletypesize{\scriptsize}
\tablecaption{Parameters for wind bubbles of early-type stars}
\tablewidth{0pt}
\tablehead{
\colhead{No.} & \colhead{type} & \colhead{$r_S$} &\colhead{exemplified star} & \colhead{$M $}
& \colhead{$\dot{M}$}& \colhead{$ \vw $}& \colhead{$\tau_{\rm ms}$}
& References & \colhead{$ \pu^{1/3}\Rb$} \\
& & $(\pc)$ & & $(\Msun)$ & $ (\Msun \yr^{-1})$ & $ (10^3\km\ps)$
& $(10^6\yr)$ & & $(\pc)$
}
\startdata
1 & B3V & 0.4 & & 8  & $1\E{-11}$ & 0.7 & 31.6 & r1,r6 & 0.9\\
2 & B2V & 0.8 & & 10 & $5\E{-10}$ & 0.7 & 22.5 & r1,r6 & 2.8 \\
3 & B1V & 1.4 & & 12 & $6\E{-9}$  & 0.7 & 16.0 & r1,r6 & 5.7\\
4 & B0.5Vp & 2.9 & HD93030 & 14 & $1.3\E{-8}$ & 0.9 & 12.6 & r2,r5,r6 & 8.4 \\
5 & B0V & 7.0 & HD149438 & 18 & $3.5\E{-8}$ & 1.2 & 9.3 & r2,r5,r6 &  12.4 \\
6 & O9.5V & 9.8 & HD38666 & 21 & $4.2\E{-8}$ & 1.1 & 7.7 & r2,r4,r5,r6 & 11.8 \\
7 & O9V & 11.7 & HD214680 & 24 & $1.3\E{-7}$ & 1.4 & 6.7 & r3,r4,r5,r6 & 19.1\\
8 & O8.5V & 13.1 & HD46149 & 27 & $1.9\E{-7}$ & 1.8 & 6.0 & r3,r4,r5,r6 & 24.5 \\
9 & O8V & 14.5 & HD101413 & 28 & $1.3\E{-7}$ & 3.0 & 5.7 & r3,r4,r5,r6 & 29.8\\
10 & O7V & 17.9 & HD47839 & 33 & $3.1\E{-7}$ & 2.6 & 5.0 & r3,r4,r5,r6 & 34.2\\
11 & O7V & 17.9 & HD35619 & 40 & $5.6\E{-7}$ & 2.5 & 4.3 & r3,r4,r5,r6 & 38.9\\
12 & O6V & 24.1 & HD101190 & 49 & $5.9\E{-7}$ & 3.2 & 3.8 & r3,r4,r5,r6 & 44.9\\
13 & O4V & 40.6 & HD46223 & 54 & $1.7\E{-6}$ & 3.1 & 3.6 & r3,r4,r5,r6 & 62.3\\
14 & O4V & 40.6 & HD242908 & 54 & $8.9\E{-7}$ & 3.2 & 3.6 & r3,r4,r5,r6 & 50.7\\
15 & O4V & 40.6 & HD164794 & 72 & $3.5\E{-6}$ & 3.4 & 3.1 & r3,r4,r5,r6 & 80.2
\enddata
\tablecomments{
The parameters of B3--B1 stars (rows 1--3),
except main-sequence age $\tau_{\rm ms}$,
are from r1 (in which the $v_w$ values are assumed).
The $M$ values of the exemplified stars of spectral types
B0.5--O9.5 (rows 4--6) and O9--O4 (rows 7--15) are adopted
from r2 and r3, respectively, and the $v_w$ values of the
exemplified stars of spectral types B0.5--B0 (rows 4--5)
and O9.5--O4 (rows 6--15) are from r2 and r4, respectively.
The $\dot{M}$ values of the exemplified B0.5--O4 stars are
obtained from r5. All of the $\tau_{\rm ms}$ values are estimated
from the evolutionary tracks of r6.
{\bf References}---r1: Chevalier (1999); r2: Snow \& Morton (1976);
r3: Bernabeu (1992);
r4: Bernabeu et al.\ (1989);
r5: de Jager et al.\ (1988);
r6: Schaller et al.\ (1992).
}
\label{T:m_rb}
\end{deluxetable*}
\end{center}

\section{Linear Relation for the Wind-Blown Bubble Size of Main-Sequence OB Stars}

\subsection{Linear Relation Expected from Theoretical Model}

According to a theoretical study of SNRs in molecular clouds by \citet{Ch99}, the maximum
size of a bubble blown by a main sequence star, when the bubble is in pressure
equilibrium with the ambient medium, is expressed as:
\begin{eqnarray} \label{eq:chev99rb}
\Rb & = &15.8\left(\frac{\dot{M}}{10^{-7}\Msun\yr^{-1}}\right)^{1/3}
  \left(\frac{\tau_{\rm ms}}{10^7 \yr}\right)^{1/3} \nonumber\\
  & & \left(\frac{v_{\rm w}}{10^3 \km\ps}\right)^{2/3}
  \left(\frac{p/k}{10^5 \cm^{-3}\K}\right)^{-1/3}
   \parsec,
\end{eqnarray}
where $\dot{M}$ and $v_{\rm w}$ are the mass-loss rate and terminal
velocity of the stellar wind, respectively, $\tau_{\rm ms}$ is the
main-sequence age,
and $p$ is the pressure of the surrounding interclump medium.
The stellar parameters can be estimated from theoretical or empirical
studies, and are ultimately functions of stellar mass mainly, as shown
below.

The mass-loss rate is dependent on the stellar luminosity, $L$, and
effective temperature, $T_{\rm eff}$ \citep{dJetal88}.  For main-sequence
massive stars, both $L$ and $T_{\rm eff}$ are functions of stellar mass, $M$;
thus, $\dot{M}$ can be expressed as a function of only $L$:
\begin{equation}
\dot{M} \propto L^{\zeta_1},
\end{equation}
where the index $\zeta_1$ has been estimated to be 1.6--1.7 by empirical
or theoretical studies.
For example, \citet{GC84} examined a sample of 50 OB stars and found
$\zeta_1 = 1.62\pm0.19$; \citet{HP89} used a sample of $\sim$200 O stars
and obtained $\zeta_1=1.69$; and \citet{Ma09} suggested $\zeta_1=1.6$
for OV stars.

The stellar luminosity is largely a function of the stellar mass, and
the mass-luminosity relation can be expressed as:
\begin{equation} L \propto M^{\zeta_2}. \end{equation}
The index $\zeta_2$ has been empirically determined by \citet{Vetal07}
to be 2.76 for a stellar mass range of 10--50 $\Msun$.
This $\zeta_2$ value is consistent with the index, 2.26--2.98,
given by Maeder (2009) for stellar masses of 9--60 $\Msun$. 

The main sequence life time, $\tau_{\rm ms}$, is proportional to the available
fuel and inversely proportional to the burning rate,
which are proportional to $M$ and $L$, respectively.
Thus,
\begin{equation}
\tau_{\rm ms}\propto \frac{M}{L} \propto M^{1-\zeta_2}. 
\end{equation}

The terminal velocity of stellar wind is proportional to the escape
velocity, and thus $v_w \propto \sqrt{M (1-\Gamma)/R_{\ast}}$,
where  $M (1-\Gamma)$ is the reduced stellar mass and $\Gamma$
is the ratio of the stellar to the Eddington luminosity.
\citet{Ma09} derived that the reduced stellar mass is a function of stellar
luminosity:
\begin{equation}
M (1-\Gamma) \propto L^{\zeta_3},
\end{equation}
with $\zeta_3=0.327$ for $10$--$120\Msun$, and that the stellar radius of
a main sequence star is a function of stellar mass:
\begin{equation} R_{\ast} \propto M^{\zeta_4}, \end{equation}
with $\zeta_4=0.56$ for 15--$120\Msun$.

Using Eqs.\ 2--6 to express $\dot M$, $\tau_{\rm ms}$, and $v_{\rm w}$
as functions of $M$ and substituting them into Eq.\ 1, we obtain
\begin{equation} p^{1/3}\Rb \propto M^{\eta}, \end{equation}
where $\eta = [\zeta_2 (\zeta_1+\zeta_3-1)-\zeta_4+1]/3$.
Adopting $\zeta_1=1.6$--$1.7$, $\zeta_2 = 2.7\pm0.3$, $\zeta_3=0.327$,
and $\zeta_4=0.56$, we find $\eta = 1.0\pm0.1$.
This result indicates that for a given interstellar pressure,
the $\Rb$--$M$ relationship is very close to linear.

\subsection{Another Derivation of the Linear Relationship}

\begin{figure}[tbh!]
\centerline{ {\hfil\hfil
\includegraphics[width=3.7in]{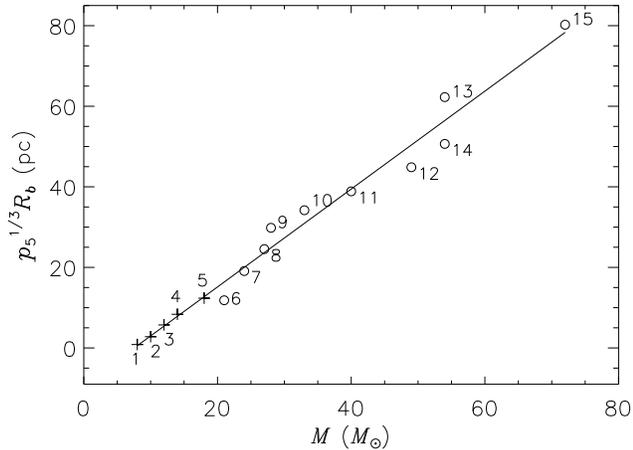}
\hfil\hfil}}
\caption{
Calculated radii of the wind-blown bubbles vs.\ the stellar masses
for OB stars.
The labeled numbers indicates various types of stars listed in
Table~\ref{T:m_rb}, with the
plus signs for B stars and the circles for O stars.
The solid line represents a linear regression,
$\pu^{1/3}\Rb=1.22M/\Msun-9.16\parsec$ [see eq.(\ref{eq:regression})].
}
\label{f:m_rb}
\end{figure}

The almost linear relationship between $p^{1/3}\Rb$ and $M$
can be derived semi-empirically using observationally determined
$\dot M$ and $v_{\rm w}$ and model-estimated $\tau_{\rm ms}$.
In Table~\ref{T:m_rb}, we have compiled 15 main-sequence stars
with spectral types ranging from B3 to O4.
For all spectral types, the $\tau_{\rm ms}$ values
are estimated from the evolutionary tracks of \citet{Setal92}.
For B3--B1 type stars, the other parameters are adopted from
\citet{Ch99}, in which the $v_w$ values are reasonably assumed.
For the exemplified B0.5--O4 stars,
the $M$ values are adopted from \citet{SM76}
and \citet{Be92}, the $v_w$ values are from \citet{SM76}
and \citet{Betal89},
and the $\dot{M}$ values are empirically given
by \citet{dJetal88}.
Using these $\dot M$, $\tau_{\rm ms}$, and $v_{\rm w}$ values in
Eq.\ 1 and defining $\pu \equiv (p/k)/(10^5\cm^{-3}\K)$, we have computed
$\pu^{1/3}\Rb$ for the 15 stars in Table~\ref{T:m_rb} and plotted
them against $M$ in Figure~\ref{f:m_rb}.
It is clear that $\pu^{1/3}\Rb$ does correlate linearly with $M$
for the interstellar wind-blown bubbles of main-sequence OB stars;
furthermore, a good linear regression for the $\pu^{1/3}\Rb$--$M$
relation can be obtained as
\begin{equation} \label{eq:regression}
\pu^{1/3}\Rb=\left[\alpha\left(\frac{M}{\Msun}\right)-\beta\right]
 \parsec,
\end{equation}
where $\alpha=1.22\pm0.05$ and $\beta=9.16\pm1.77$.
If the intercloud pressure $p/k$ is constant and $\approx 10^5 \cm^{-3}\K$
(i.e., $\pu\approx1$) as suggested, e.g., by Chevalier's (1999), then
$\Rb$ is linearly correlated with $M$.

\subsection{Post-Main Sequence Evolution} \label{S:applicability}
Whether a wind-blown bubble may survive the post-main sequence evolution
without much change until the SN explosion depends on the initial
mass of the central massive star.
In a recent review, \citet{Sm09} concludes that RSGs with an initial mass
in the range 8 to 25--$30\Msun$ can be progenitors for SNe~II-P and those
with an initial mass above $17\Msun$ could end as SNe~II-L, IIn, and Ib/c.
In the He-core burning stage, the
RSG wind expands in the low-density interior of the
main-sequence bubble with the outer edge of the RSG wind bounded by the
pressure of the surrounding gas. 
For RSGs terminating their lives in SNe~II-P, their winds reach an
outer radius of $\la 1\parsec$, and for RSGs ending in SNe~II-L their
winds' outer boundaries approach $\sim 5\parsec$ \citep{Che05}.
These radii are
much smaller than the expected $R_{\rm b}$; thus their
RSG winds do not interact with the shell of the main-sequence bubble at all.

For an initial mass greater than 25--$30\Msun$, however, the star will
evolve through an RSG phase or an LBV phase before turning into a
WR star \citep[e.g.,][]{Ma09,Sm09}.
The fast, strong WR wind will interact with the circumstellar material shed by the
star during the RSG or LBV phase and form a circumstellar bubble.
Observations have shown that the radii of circumstellar bubbles
are small, usually $\le$1 pc for LBV bubbles and a few pc for WR bubbles
\citep{Chu03}, much smaller than their corresponding main-sequence
interstellar bubbles; furthermore, infrared (IR) observations of WR bubbles have
detected larger shell structures that correspond to the main-sequence
interstellar bubbles \citep[e.g.,][]{NF93,Ma96}.
Nevertheless, it is possible that a WR star's circumstellar bubble has
merged with its main-sequence progenitor's interstellar bubble, in which
case no strong abundance anomaly is seen and the bubble expansion velocity
is only $\sim$20 km\,s$^{-1}$, such as NGC\,2359 and NGC\,3199 \citep{Esetal92}.
These merged bubbles may be further accelerated by the WR winds.
At the time of SN explossion, the bubble size may or may not have grown
significantly from the stalled main-sequence interstellar bubble.


\begin{center}
\begin{deluxetable*}{ccccc|c}
\tabletypesize{\scriptsize}
\tablecaption{Galactic SNRs with molecular shells/in molecular cavities}
\tablewidth{0pt}
\tablehead{
\colhead{SNR} & \colhead{attribute$^{\rm a}$} & \colhead{radius of bubble}
   & \colhead{distance}& \colhead{Reference$^{\rm b}$} & \colhead{progenitor mass}\\
   & & (pc) & (kpc) & & ($\Msun$)
}
\startdata
G21.8$-$0.6 (Kes~69) & S & 13 & 6.2 & r1 & $18\pm2$ \\
G29.7$-$0.3 (Kes~75) & S & 6 & 10.6 & r2 & $12\pm2$ \\
G32.8$-$0.1 (Kes~78) & S & 17 & 4.8 & r3 & $21\pm2$ \\
G39.2$-$0.3 (3C396) & C & 7 & 6.2 & r4 & $13\pm2$ \\
G41.1$-$0.3 (3C397) & C & 4.5--7 & 10.3 & r5 & $12\pm2$ \\
G54.4$-$0.3 (HC~40) & S  & 18/43 & 3/7 & r6 & 22$\pm2$/? \\
G263.9$-$3.3 (Vela) & S & 14--19 & 0.29 & r7, r8 & $21\pm3$ \\
G347.3$-$0.5 (RXJ1713$-$3946) & C & 9 & 1.1 & r9, r10 & $15\pm2$
\enddata
  \tablenotetext{a}{\phantom{0}C --- SNR is discovered to be in a molecular cavity;
  S --- SNR has a coincident molecular shell-like structure.}
  \tablenotetext{b}{\phantom{0}References---r1: Zhou et al.\ (2009);
r2: Su et al.\ (2009); r3: Zhou \& Chen (2011);
r4: Su et al.\ (2011); r5: Jiang et al.\ (2010);
r6: Junkes et al.\ (1992); r7: Moriguchi et al.\ (2001);
r8: Dodson et al.\ (2003);
r9: Fukui et al.\ (2003); r10: Moriguchi et al.\ (2005). }
\label{T:progenitor}
\end{deluxetable*}
\end{center}

\section{Application to SN progenitors}




Among the core-collapse SNe, more than $\sim$75\% are of Type II
and the greatest majority of these
are of Type II-P \citep{Sm09}, for which the SN explosions occur
during the RSG phase and the initial masses of the progenitors
fall in the range of 8 to 25--30$\Msun$. 
For this mass range, if a progenitor star was in a
molecular environment with $\pu^{1/3}\approx1$, its SN will
explode within a bubble that follows the linear $\Rb$--$M$ relation
described in Section 2.
Therefore, the size of such a SNR may reflect the size of the
main-sequence interstellar bubble and the $\Rb$--$M$ relation can
be used to estimate the initial mass of the SN progenitor.

SNe Ib/c have been suggested to originate from WR progenitors that
had shed most or all of their H envelopes \citep{Getal86} or
moderate-mass interacting binaries in which the exploding star's envelope
had been stripped through Roche lobe overflow or common-envelope
evolution \citep{Petal92,Netal95}.  According to population analysis, it is
concluded that WR stars cannot be responsible for all SNe Ib/c and that
binary production of SNe Ib/c has to be significant \citep{Sm09}.
Based on theoretical modeling, it is suggested that massive WR stars
may collapse into black holes without SN explosions
\citep[e.g.,][]{F99,Hetal03}.
In view of these factors, the sizes of SN Ib/c remnants cannot
be used to estimate the stellar progenitors' masses, as it is uncertain
whether and how the binary progenitors form interstellar bubbles and it
is also uncertain to what extent the WR winds affect the evolution of
the main-sequence bubbles.


Multiwavelength observations of Galactic SNRs have shown that at least
a few tens of them are evolving in a molecular cloud environment
\citep{Jetal10}.  Some of these SNRs appear to be located within
molecular gas cavities or in contact with molecular shells, which
are likely the relics of wind-blown bubbles created by the stellar
progenitors.
The photoionized regions are significantly smaller than the
wind-blown bubbles in size for massive stars.
A comparison in size between the wind-blown bubbles and photoionized
regions can also be seen in Table~\ref{T:m_rb}, where the
``Str\"{o}mgren radii" (see Spitzer 1978) $r_S$ are calculated for
an interclump density $5$~H~atoms$\cm^{-3}$ (Chevalier 1999) and
the ionizing flux for each spectral type is adopted from Panagia (1973).
Assuming that these cavities or shells are indeed interstellar
bubbles blown by the progenitors during the main-sequence stage
and the bubbles had been stalled in pressure equilibrium with the
ambient interclump medium at $\pu\approx1$, we may use the bubble
sizes and Eq.\ 8 to estimate the masses of their SN progenitors.
Below, we discuss eight SNRs and carry out the mass estimates individually.
The results are summarized in Table~\ref{T:progenitor}.

{\bf G21.8$-$0.6 (a.k.a.\ Kes\,69)}.
In this SNR, a partial molecular arc at a systemic velocity of
$v_{\rm LSR} \sim85\km\ps$
is detected in the southwest with morphological correspondence to
the brightened partial SNR shell seen in the radio, IR and X-ray
wavelengths; it is suggested to be a part of the cooled material swept
up by the progenitor's stellar wind \citep{Zetal09}.  Assuming that
the molecular arc is associated with the main-sequence bubble, its 13~pc radius
implies that the SN progenitor star had an initial mass of $\sim18\pm2\Msun$,
of spectral type around B0.

{\bf G29.7$-$0.3 (a.k.a.\ Kes\,75)}.
This SNR is in a cavity surrounded by a molecular shell unveiled
in the broadened blue wing of the $^{12}$CO line at
$v_{\rm LSR} \sim 54\km\ps$, and the southern part of the molecular
shell is likely the cooled, clumpy shell of the progenitor's main-sequence
bubble \citep{Setal09}.
Thus, the progenitor of Kes\,75 may have an initial mass $\sim12\pm2\Msun$,
corresponding to a B0.5--B2 star, which is expected to end its life in a Type~II-P
SN explosion \citep{Hetal03, Sm09}.
The pulsar PSR J1846$-$0258 at the center of Kes\,75 has been suggested to
be a magnetar \citep{Gavriil08, Kumar08}.
Although a few magnetars are inferred to have progenitors with mass $>30\Msun$,
smaller mass is also suggested (e.g., $17\Msun$ for SGR 1900+14),
and the mass of the magnetar's progenitor may span a wide range
\citep[see][and references therein]{SK12}.

{\bf G32.8$-$0.1 (a.k.a.\ Kes\,78)}.
Kes\,78 is found to be interacting with a clumpy molecular cloud, which
is at a systemic velocity of $v_{\rm LSR} \sim 81\km\ps$ and shows a
clumpy arc in the west bearing kinematic signatures of shock perturbations
\citep{ZC11}.
The entire SNR appears to be in a cavity surrounded by molecular material.
If the molecular cavity of Kes\,78, 17~pc in radius, was created by the
progenitor's main-sequence wind, our linear relation implies a progenitor
mass of $\sim 21\pm2\Msun$, corresponding to a spectral type around O9.5.

{\bf G39.2$-$0.3 (a.k.a.\ 3C\,396)}.
3C\,396 is coincident with a molecular cavity at $v_{\rm LSR} \sim
85$--$87\km\ps$, and the western boundary of the SNR perfectly
follows the eastern face of a molecular wall \citep{Setal11}.
\citet{Letal09} attribute the near-IR [Fe II] line emission
to the SNR shock overtaking the RSG wind of the SN progenitor,
and suggest a progenitor mass of 25--$35\Msun$. Such a mass
is higher than the mass limit, $\la25\Msun$, for neutron stars'
progenitors with solar metallicity \citep{Hetal03}.
Assuming that the abundances of the X-ray-emitting gas in 3C\,396
are enriched by the SN ejecta, \citet{Setal11} derived a progenitor
mass of 13--$15\Msun$, suggesting a spectral type of B1--B2.
If we regard the cavity as the extent of the progenitor's main-sequence
bubble, then the progenitor's mass estimated from our linear relation
is $\sim13\pm2\Msun$, consistent with that estimated from
the ejecta abundances.

{\bf G41.1$-$0.3 (a.k.a.\ 3C\,397)}.
3C\,397 is a radio- and X-ray-bright Galactic SNR with a peculiar rectangular
morphology \citep{Cetal99,SHetal05} and its Fe-rich ejecta essentially
aligning along a diagonal of the rectangle \citep{JC10}.
The SNR is confined in a molecular cavity with an extent $9\times14$~pc$^2$
at $v_{\rm LSR}\sim32\km\ps$, and the cavity was likely sculpted chiefly
by the progenitor star and the cavity walls hampered the expansion of the
ejecta, sending a reflected shock back to the ejecta \citep{Jetal10}.
Thus, using the size of this molecular cavity we estimate the progenitor
mass to be $\sim12\pm2\Msun$, which corresponds to a spectral type B0.5--B2.
A recent \XMM-Newton X-ray study of this
remnant has analyzed the metal abundances of the SN ejecta and thus
given an independent assessment of the progenitor's mass, $11$--$15\Msun$
(Safi-Harb et al.\ in preparation), in good agreement with our estimate.

{\bf G54.4$-$0.3 (a.k.a.\ HC\,40)}.
SNR~HC\,40 exhibits a perfect CO shell at $v_{\rm LSR} \sim$ 36--$44\km\ps$
aligning with the SNR's radio continuum shell \citep{Jetal92}.
This radial velocity corresponds to two kinematic distances, 3 and
7~kpc, but the shorter distance is favored because it agrees with
an independent distance estimate for nearby HII regions and
OB association.
Using the 3 kpc distance and assuming that the CO shell corresponds to
the main-sequence bubble, we estimate an initial mass of $\sim22\pm2\Msun$
for the progenitor, which is in the range for Type~II SN explosion
of a RSG \citep{Hetal03,Sm09}.  If the 7 kpc distance is adopted,
the CO shell radius will be larger than 30 pc and the progenitor mass
would be so high ($>30\Msun$) that it may be expected to collapse into a
black hole without an energetic SN explosion.
Thus, we suggest that HC\,40 is at a distance of $\sim$3 kpc and that
its progenitor was an B0--O9 star.


%
{\bf G263.9$-$3.3 (a.k.a.\ Vela)}.
The Vela SNR has been shown to coincide with a molecular void of angular
diameter 5.4--7.4$^{\circ}$, delineated by molecular clumps in a velocity
range of $v_{\rm LSR}$ =  $-5$ to $85\km\ps$,
with a total mass of the order of $10^{4}\Msun$.
It is further suggested that the molecular clumps are
pre-existent, rather than having been swept up by the SNR shock, and
that the SNR may have been expanding in a low density
(of the order of $10^{-2}\cm^{-3}$) medium \citep{Metal01}.
Thus, it is reasonable to assume that the low-density region is a
wind-blown cavity enclosed by the observed molecular clumps.
A pre-existent wind-driven shell, currently  impacted by the SN
ejecta/shock, has been previously proposed to explain the filamentary
structures seen in radio, optical, and X-ray bands \citep{Gv99}.
The VLBI parallax measurements suggest a distance of 290 pc
\citep{Detal03}.
For this distance, the cavity has a radius of 14--$19\pc$, which then
leads to an inference of a B0--O9 progenitor star,
with an initial mass of $21\pm3\Msun$, if it was a single star.
This estimate similar to a previous suggestion of a 15--$20\Msun$
progenitor for a Type II-P SN explosion based on the moderate size
of the wind-blown bubble \citep{Gv99}.



{\bf G347.3$-$0.5 (a.k.a.\ RX~J1713.7$-$3946)}.
This SNR is a TeV $\gamma$-ray and non-thermal X-ray source, and it
appears to be confined in a molecular gas cavity at $v_{\rm LSR}$
$\sim-11$ to $-3\km\ps$ \citep{Fetal03,Metal05}.
It is suggested that this SNR is still in the free expansion phase
and the non-decelerated blast wave is colliding with the dense molecular
gas after it traveled in a low-density cavity that perhaps was produced
by the stellar wind or pre-existing SNe \citep{Fetal03}.
If we assume the cavity was created by the wind of a single
progenitor star, then the star might has an initial mass of
$\sim15\pm2\Msun$, with a spectral type B0--B1.

It would be safe, however, to regard the above mass estimates as
lower limits (e.g., in the case of Kes~69).
Actually, in some cases, bubble's blowout from the side of a molecular
cloud can be possible, and thus the progenitors' masses may be underestimated.
But because $\Rb\propto p^{-1/3}$ and the dependence on $p$ is insensitive,
the underestimate would be small, if most volume of the bubble is in the cloud.

\section{Summary}

For OB stars in a molecular gas environment with a postulated constant
interclump pressure $p/k\sim10^5 \cm^{-3}\K$, we find a linear relation
between the sizes of the main-sequence wind-blown bubbles and the stellar
masses: $\Rb\approx1.22M/\Msun-9.16\parsec$.
Since stars in the mass range 8 to 25--$30\Msun$ will end their lives
in the RSG phase and will not launch a further WR wind,
the extent of molecular gas cavities are largely determined by
the main-sequence wind bubbles.
The linear $\Rb$--$M$ relation can thus be used to assess the initial
masses of SN progenitors for SNRs evolving in molecular cavities.
We have applied this method to estimate the masses of the stellar progenitors
of eight SNRs: Kes~69, Kes~75, Kes~78, 3C~396, 3C~397, HC~40, Vela,
and RX~J1713$-$3946. The progenitor masses of these eight SNRs are in the
range of 10--24 $M_\odot$.

\begin{acknowledgements}
We thank Lida Oskinova for advice on the physical parameters of B stars.
Y.C.\ thanks Xinlian Luo and Bing Jiang for helpful discussion on stellar evolution
and relevant SNRs, respectively.
Y.C.\ acknowledges the support from NSFC grant 11233001,
the 973 Program grant 2009CB824800,
grant 20120091110048 by the Educational Ministry of China,
and the grants by the 985 Project of NJU and the Advanced
Discipline Construction Project of Jiangsu Province.
Y.H.C.\ acknowledges the support of SAO/CXC grant GO0-11025X.
\end{acknowledgements}

\end{document}